# Cosmic dust as a prerequisite for the formation of complex organic molecules in space?

Alexey Potapov[1*], Kilian Pollok[1], Falko Langenhorst[1], Martin McCoustra[2], Robin T. Garrod[3]

[1]*Analytical Mineralogy Group, Institute of Geosciences, Friedrich Schiller University Jena, Jena, Germany*

[2]*Institute of Chemical Sciences, Heriot-Watt University, Edinburgh, UK*

[3]*Departments of Astronomy & Chemistry, University of Virginia, Charlottesville, VA, USA*

**Abstract**

In cold, dense astrophysical environments dust grains are mixed with molecular ices. Chemistry in those dust/ice mixtures is determined by diffusion and reaction of molecules and radicals. However, investigations of diffusion of astrophysically relevant radicals and molecules across the surface and through the pores of cosmic dust grains and of surface reactions consequent to such diffusion is largely uncharted territory. This paper presents results of a study of a solid-state reaction of two molecular species, $CO_2$ and $NH_3$, separated by a layer of porous silicate grain aggregates, analogues of cosmic dust. The experiments demonstrate that the presence of the dust layer was necessary for a pure thermal $CO_2 + 2NH_3$ reaction to proceed, leading to the formation of ammonium carbamate ($NH_4^+NH_2COO^-$), an ionic solid containing a complex organic moiety of prebiotic interest recently detected in a protoplanetary disk. This result speaks for: (i) efficient diffusion of molecules on/within cosmic dust, (ii) an underestimated role for surface catalysis in the astrochemistry of cosmic dust, and (iii) potentially efficient dust-promoted chemistry in warm cosmic environments, such as protostellar envelopes and protoplanetary disks.

*Corresponding author, email: alexey.potapov@uni-jena.de

# 1. Introduction

Surface reaction pathways lead to a greater complexity of molecular species (as compared to the gas phase) in cold astrophysical environments, such as interstellar clouds, protostellar envelopes and protoplanetary disks beyond the water snowline. The surface is provided by dust grains, mainly based on carbonaceous and siliceous materials, such as amorphous and crystalline silicates, amorphous carbon, graphite, polycyclic aromatic hydrocarbons, and fullerenes. In addition to providing a meeting place for reactants, the dust serves as a third body and a possible catalyst for reactions thereon (Potapov & McCoustra 2021).

There are different models/views of cosmic dust in cold astrophysical environments. The current most popular view describes a dust particle as a compact refractory dust core covered by a thick ice layer of tens or hundreds of monolayers (Ehrenfreund et al. 1998; Allamandola et al. 1999; Burke & Brown 2010). The "onion" model (Ehrenfreund, et al. 1998) visualizes the icy coat of the grains as a series of layers, where the core dust grain is first fully covered by a thick layer of water-rich ice (water is the main constituent of ices in cold and dense regions). This view practically excludes the dust surface from physical and chemical processes thereon, which should occur on/in thick ices covering the dust.

However, there is considerable evidence in support of high porosity and surface area of interstellar and circumstellar dust grains, from laboratory experiments on cosmic dust analogues, dust evolution models, astronomical observations, and analysis of cometary and interplanetary dust particles. The interested reader is referred to (Potapov et al. 2020a), where a new view of cosmic dust grains was presented, and to a recent review on the porosity of cosmic dust (Potapov et al. 2025b). In (Potapov, et al. 2020a), it was demonstrated that the coverage of porous grains by water ice could typically be in the range from the sub-monolayer to a few monolayers. Thus, in this scenario, the surface of the dust itself could remain available for physical and chemical processes thereon. The same conclusion was reached in (Marchione et al. 2019), where it was shown that - due to the self-aggregation of $H_2O$ molecules – the bare dust grain surface in cold astrophysical environments would be available for the adsorption of species other than water directly, even after a substantial ice build-up. Moreover, the first evidence of dust/ice mixing in protostellar envelopes and protoplanetary disks was provided a few years ago by comparison of observational and laboratory spectra (Potapov et al. 2021), supporting the new view.

There are three common types of reactions at a surface: Eley-Rideal reactions, Langmuir–Hinshelwood reactions, and Kasemo-Harris (or hot atom) reactions (see, e.g., (Potapov & McCoustra 2021)). The first describes reactions between a gaseous species and a species adsorbed on the surface at a single site, and the latter two types consider diffusion of at least one of the reactants as a prerequisite for the reaction. Thus, reaction and diffusion are simultaneously competing and complementing processes. However, diffusion on astrochemically relevant surfaces is poorly explored and was mainly focused on diffusion in/on molecular ices (e.g., (Congiu et al. 2014; Kouchi et al. 2020; Tsuge et al. 2023) and references therein).

The goal of the present study was to explore the pure thermal reaction $CO_2 + 2NH_3 \rightarrow NH_4^+NH_2COO^-$. The product, ammonium carbamate, is an ionic solid and a salt containing a complex organic anion of prebiotic interest as it is a direct precursor of urea. The molecule was for a long time expected to be present in astrophysical environments and was recently detected in a protoplanetary disk (Potapov et al. 2025a). The $CO_2 + 2NH_3 \rightarrow NH_4^+NH_2COO^-$ reaction has been intensively studied in mixed $CO_2/NH_3$ bulk ices (Bossa et al. 2008; Noble et al. 2014; Ghesquiere et al. 2018) and in mixed ices on the surface of cosmic dust grain analogues (Potapov et al. 2019; Potapov et al. 2020b), in the latter case demonstrating the catalytic effect of the dust.

However, the reaction in mixed ices proceeds efficiently only with an excess of $NH_3$ (Noble, et al. 2014); such a scenario would appear unlikely in a realistic astrophysical environment considering the much lower abundance of $NH_3$ as compared to $CO_2$ in interstellar and circumstellar ices (Boogert et al. 2015). Thus, despite extensive laboratory investigations, the formation route to ammonium carbamate in astrophysical environments is still an open question.

This study presents a unique case – the $CO_2 + 2NH_3 \rightarrow NH_4^+NH_2COO^-$ reaction in the presence of a dust layer separating the $CO_2$ and $NH_3$ reactants. Thus, diffusion of reactants is unavoidable for the reaction to proceed. The questions, which the authors wanted to answer were: will the reaction proceed in principle (whether the molecules have ability to diffuse through the dust) and, if yes, what is the influence of the dust layer thickness on the reaction yield and rate and whether diffusion through dust may clarify the formation route to ammonium carbamate?

There is a principal difference between the previous studies of the formation of ammonium carbamate on the surface of cosmic dust grain analogues (Potapov, et al. 2019; Potapov, et al. 2020b) and the present one. In the previous studies, the reactants, $CO_2$ and $NH_3$, were premixed on the surface before the reaction proceeded. Thus, it was not possible to differentiate between the case of reaction without diffusion and the case of reaction with diffusion and conclude, whether diffusion plays any role. In the present study, the reactants are separated by a layer of dust and the reaction cannot physically proceed without diffusion. Thus, diffusion through the dust is a prerequisite for the reaction. The results presented below are the first evidence of efficient diffusion and reaction of molecules on and through highly porous dust aggregates, analogues of cosmic dust grains, leading to the formation of a complex organic molecule.

2. Methods

The experiments were performed in the Jena Dust Machine allowing for simultaneous deposition of analogues of cosmic dust particles and molecular ices (see, e.g., (Potapov et al. 2023; Potapov et al. 2024) and references therein). Amorphous magnesium silicate ($MgSiO_3$ stoichiometry) grains were used as analogues of cosmic silicates. The structure and morphology of the dust layers were analyzed using the combination of Focused Ion Beam (FIB) preparation and transmission electron microscopy (TEM). The surface on the $MgSiO_3$ sample was coated by a thin layer of gold to make it conductive and to protect the surface from damage during the first ion imaging. FIB preparation was performed using a dual beam FIB-SEM FEI Quanta 3D FEG operating with a Ga ion gun at 30 kV and beam currents between 30 to 0.3 nA. A deposited Pt stripe was used to protect the FIB cross section during the ion thinning process. The FIB section was finally milled to a thickness of about 100 nm. TEM imaging was done conventionally (mass thickness contrast, bright field and high-resolution imaging) and by low magnification scanning transmission electron microscopy (LM-STEM) using a 200 kV FEI Tecnai G2 F20 TEM. LM-STEM allows to image the complete FIB cross section to measure the mean thickness of the layer over a length of 21 μm. The composition of the deposited layer was measured by energy-dispersive X-ray (EDX) emission spectroscopy using an Oxford X-MaxN 80T SDD system which confirmed the $MgSiO_3$ stoichiometry of the deposited layer.

The FIB-TEM analysis of the deposited layer with material for nominally 100 nm thickness shows a rough surface with an average layer thickness of 666 nm with thickness variation of ± 185 nm (Fig. 1a). The layer consists of small (up to 10 nm) individual amorphous particles which merged into a foamy structure with pore diameters ranging dominantly from 10 to 50 nm and few pores above 100 nm (Fig. 1b&c). Thus, the layer possesses an interconnected porosity of 80-90 % resulting in a very high surface area and providing diffusion channels for reactants.

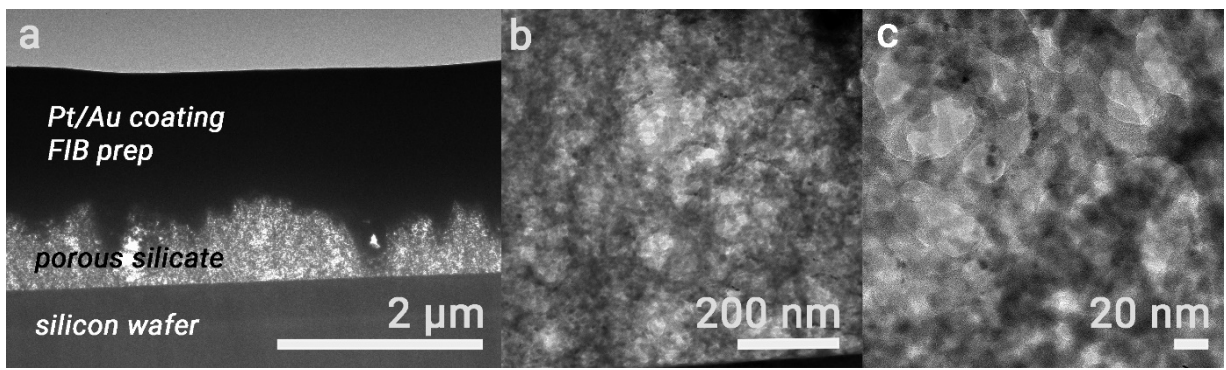

Figure 1. a) TEM image (mass thickness contrast) of a FIB prepared cross section showing a deposited porous magnesium silicate layer on a silicon wafer. The layer has a variable thickness and a very high porosity. The thickness of the prepared TEM sample in beam direction is about 100 nm. The metal coating is protecting the surface during the focused ion beam sample preparation. b) and c) Magnified view of the porous and foam-like structure of the film. Brighter areas represent less solid material in the beam direction. Areas with pore diameters over 100 nm are scarce, most pores measure less than 20 nm.

Three types of sandwich-like samples were deposited at 10 K: $CO_2$-ice/$NH_3$-ice, $CO_2$-ice/$H_2O$-ice/$NH_3$-ice, and $CO_2$-ice/$MgSiO_3$-grains/$NH_3$-ice (the sequence of species indicates the deposition sequence) with $CO_2$ and $NH_3$ ice thicknesses of 40 and 180 nm and various $MgSiO_3$-grains thicknesses from 10 to 210 nm. Note that these are nominal thicknesses, not considering a very high porosity of the dust layers. Thus, at the initial conditions two reactants, $CO_2$ and $NH_3$, were completely separated. After the deposition, the temperature of the samples was increased to 80 K and isothermal kinetic experiments were performed for 4 hours monitoring the formation of ammonium carbamate ($NH_4^+NH_2COO^-$), the product of pure thermal reactions between $CO_2$ and $NH_3$, following our previous study (Potapov, et al. 2019). Figure 2 presents examples of the deposition spectra for the case of 50 nm deposition of $MgSiO_3$. The species responsible for the main vibrational bands are assigned in the figure. Note that $^{13}CO_2$ was used.

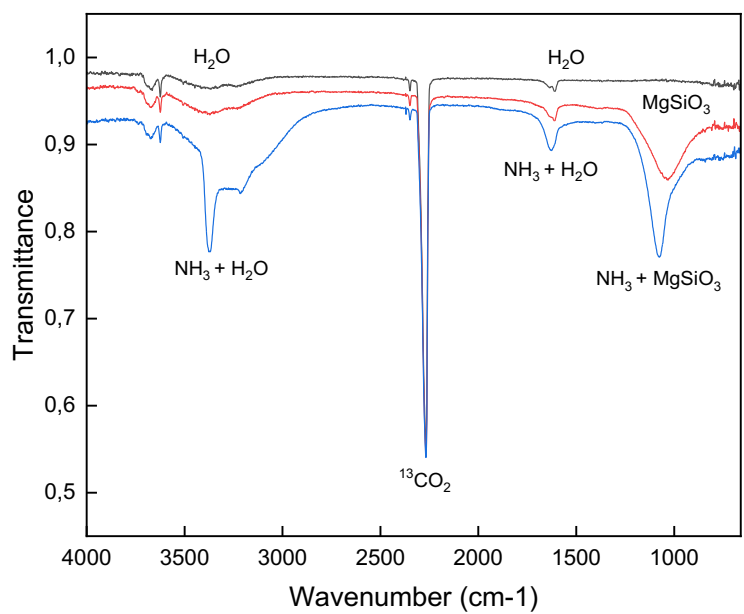

Figure 2. IR spectra for the case of 50 nm deposition of MgSiO$_3$: upper – after CO$_2$ deposition, middle – after MgSiO$_3$ deposition, lower – after NH$_3$ deposition. The spectra are shifted vertically for clarity. The species responsible for *the* main vibrational bands are assigned in the figure. The small band on the left side of the $^{13}$CO$_2$ band is $^{12}$CO$_2$ contamination in the amount of less than 1% with respect to $^{13}$CO$_2$.

The experiments were performed at pressures of about 9×10$^{-8}$ (10 K) and 2×10$^{-7}$ (80 K) mbar. Under such vacuum conditions, it is not possible to avoid accretion of water from the chamber volume. Thus, (i) all layers were mixed with water ice and (ii) water accreted on top of the samples during the isothermal kinetic experiments. Top accretion on the NH$_3$ ice should not influence the motion of NH$_3$ molecules at lower layers. Thus, the second factor can be considered unimportant for the results and conclusion presented. Regarding the first factor, the accretion rate of water at 10 K was measured to be 0.8 nm min$^{-1}$, which is lower than can be expected for the given pressure. This could result from the difference between the total pressure and the partial pressure of water in the vacuum chamber. CO$_2$ and NH$_3$ were deposited during 10 minutes in amounts of 40 and 180 nm, thus, giving the ratios CO$_2$:H$_2$O 5:1 and NH$_3$:H$_2$O 22:1. MgSiO$_3$ layers were deposited with the rate of 7-8 nm min$^{-1}$, thus, the amount of accreted water was proportional to the dust thickness with the ratio MgSiO$_3$:H$_2$O of about 10:1. Thus, CO$_2$, NH$_3$, and MgSiO$_3$ dominated in the samples. The presence of water ice mixed with CO$_2$ and NH$_3$ ices and dust grains can be even seen positively, making the samples more realistic as relevant to cold astrophysical environments. In the $CO_2$-ice/$H_2O$-ice/$NH_3$-ice sample, the thickness of the water layer was about 30 nm.

The $H_2O$, $CO_2$, and $NH_3$ ice thicknesses were calculated from their vibrational bands at 3250, 2280, and 1073 cm$^{-1}$ using the band strengths of $2\times10^{-16}$ cm molecule$^{-1}$ (Hudgins et al. 1993), $7.8\times10^{-17}$ cm molecule$^{-1}$ (Gerakines et al. 1995), and $1.7\times10^{-17}$ cm molecule$^{-1}$ (d'Hendecourt & Allamandola 1986) correspondingly and assuming a monolayer column density of $1\times10^{15}$ molecules cm$^{-2}$ and a monolayer thickness of 0.3 nm. The column density of $NH_4^+NH_2COO^-$ was calculated from the 1522 cm$^{-1}$ band using the band strength of $2\times10^{-17}$ cm molecule$^{-1}$ (Bossa, et al. 2008).

## 3. Results

Ammonium carbamate was detected in all samples containing a dust layer. Figure 3 presents an example of the spectra before and after the isothermal kinetic experiment at 80 K. Two bands related to $NH_4^+NH_2COO^-$ (**$^{13}C$**) at 1522 and 1358 cm$^{-1}$ were detected. Note that the usage of $^{13}CO_2$ led to shifts of the band positions as compared to the literature values (e.g., (Potapov, et al. 2019)). Thus, the reaction $CO_2 + 2NH_3 \rightarrow NH_4^+NH_2COO^-$ successfully proceeded in the samples. Top accretion of $H_2O$ on the $NH_3$ ice due to the vacuum conditions was not avoidable and is clearly seen in the spectrum taken after the isothermal kinetic experiment.

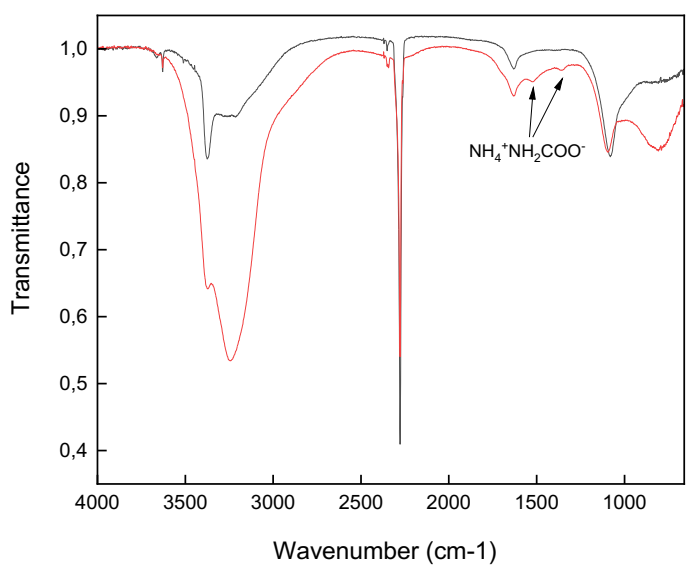

Figure 3. IR spectra before and after 4 hours of the isothermal kinetic experiment at 80 K for the sample containing a 50-nm $MgSiO_3$ layer. Two bands related to $NH_4^+NH_2COO^-$ ($^{13}C$) are shown by arrows.

Figures 4 and 5 show how the ammonium carbamate bands evolve with dust layer thickness, and reaction time at 80 K.

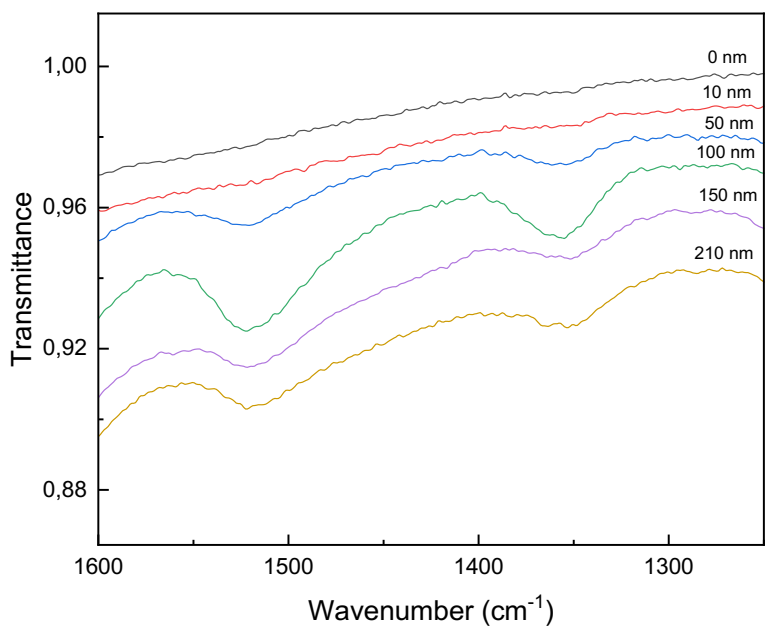

Figure 4. Evolution of the ammonium carbamate bands with the dust layer thickness after 4 hours of the isothermal kinetic experiment at 80 K. The spectra are shifted vertically for clarity.

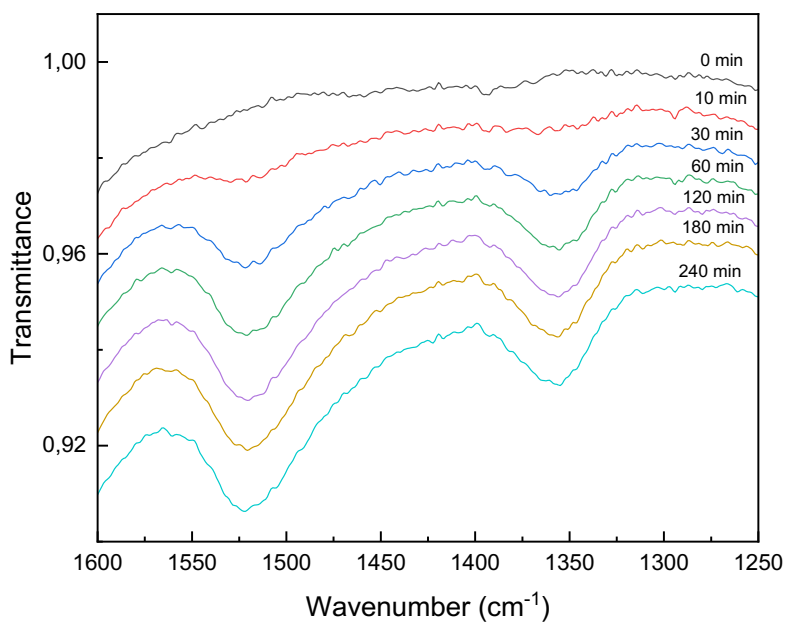

Figure 5. Evolution of the ammonium carbamate bands with the reaction time at 80 K for the sample containing a 100-nm $MgSiO_3$ layer. The spectra are shifted vertically for clarity.

The results of the isothermal kinetics experiments at 80 K are shown in Figure 6. No ammonium carbamate was detected in the $CO_2$-ice/$NH_3$-ice and $CO_2$-ice/$H_2O$-ice/$NH_3$-ice samples. In case of ammonium carbamate formation, the exponential decay of $CO_2$ and the corresponding growth of $NH_4^+NH_2COO^-$ exhibit pseudo first-order reaction kinetics (Noble, et al. 2014). The reaction rate coefficients were obtained from the isothermal curves following the formation of ammonium carbamate as in our previous studies (Potapov, et al. 2019; Potapov, et al. 2020b). The reaction rate coefficients for the samples studied are presented in Table 1 together with the reaction yields (i.e. the column densities of $NH_4^+NH_2COO^-$ at the end of the isothermal kinetic experiments). In total, about $1.2\times10^{17}$ $CO_2$ molecules were deposited in one experiment. Thus, we can conclude that maximal reaction yield corresponds to about 50% of $CO_2$ molecules converted into ammonium carbamate.

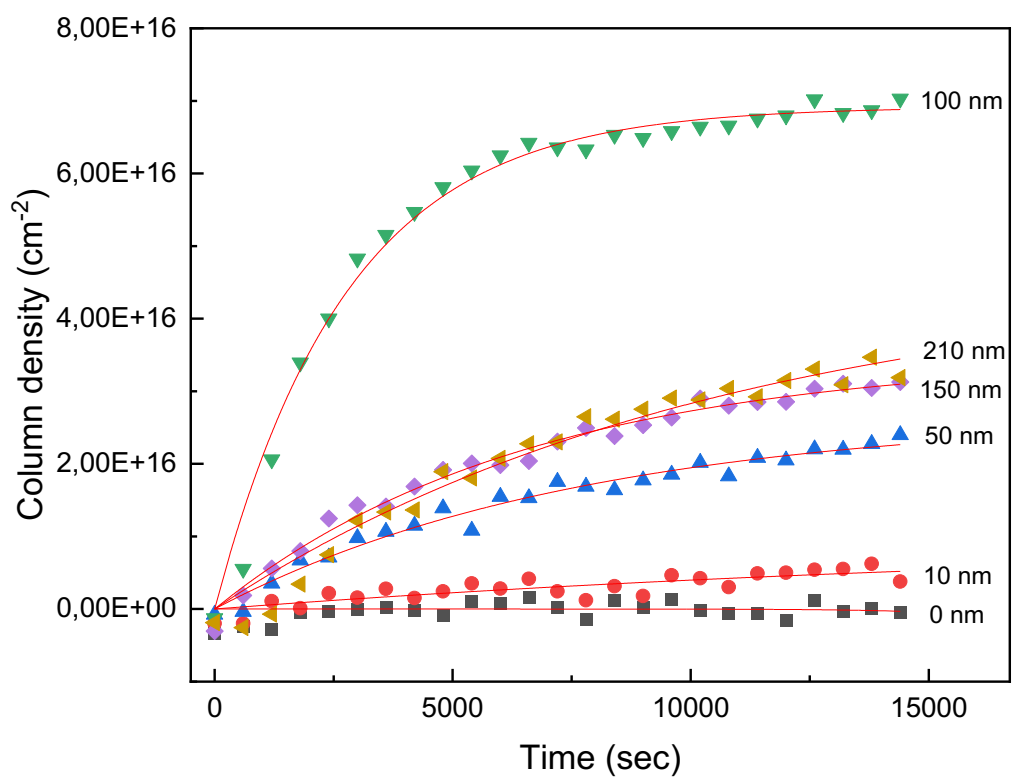

Figure 6. Time dependences of the $NH_4^+NH_2COO^-$ column density calculated from the 1522 $cm^{-1}$ band for all samples except $CO_2$-ice/$H_2O$-ice/$NH_3$-ice and their fits. The nm values are the dust layer thicknesses. Errors are reflected by the variation of the signal for 0 nm around the 0-line.

Table 1. Reaction rate coefficients (in $10^{-4}$ s$^{-1}$) and reaction yields (in $10^{16}$ cm$^{-2}$) of the CO$_2$ + 2NH$_3$ → NH$_4^+$NH$_2$COO$^-$ reaction at 80 K for different dust layer thicknesses.

| Dust layer thickness, nm | Reaction rate coefficient | Reaction yield |
|---|---|---|
| 0, H$_2$O* | 0 | 0 |
| 0 | 0 | 0 |
| 10 | 0.5 ± 0.4 | 0.5 ± 0.4 |
| 50 | 1.3 ± 0.1 | 2.3 ± 0.3 |
| 100 | 3.6 ± 0.2 | 6.9 ± 0.6 |
| 150 | 1.5 ± 0.1 | 3.1 ± 0.3 |
| 210 | 0.9 ± 0.2 | 3.4 ± 0.6 |

*CO$_2$-ice/H$_2$O-ice/NH$_3$-ice sample

## 4. Discussion

**Diffusion through dust**

The first important result is the absence of the NH$_4^+$NH$_2$COO$^-$ signal in the CO$_2$-ice/NH$_3$-ice case. This result attests for no or very slow diffusion of CO$_2$ into NH$_3$ and vice versa even at a relatively high temperature of 80 K relevant to warm environments, such as protostellar envelopes and protoplanetary disks. Separation of the reactants by water ice (which could play a catalytic role) does not improve the situation, however, this result is not surprising as it was shown that efficient diffusion of CO$_2$ and NH$_3$ molecules through water ice takes place only at temperatures above 100 K, driven by self-diffusion of water molecules in the ice (Mispelaer et al. 2013; Ghesquiere et al. 2015).

Detection of ammonium carbamate in the CO$_2$-ice/MgSiO$_3$-grains/NH$_3$-ice samples is a clear sign of efficient diffusion of reactants (as we excluded their diffusion into each other) through or, more precisely, through the pores of the dust. The increase of the reaction rate coefficient and yield with the increase of the dust layer thickness from 10 to 100 nm can be clearly explained by the increasing dust surface area and correspondingly increasing number of binding sites, where CO$_2$ and NH$_3$ molecules can meet and react. The negative slope of the rate/thickness dependence after 100 nm can be explained by increasing time required for the reactants to diffuse and by partial blocking of diffusion channels in lower dust layers by upper ones. The rate coefficients presented in Table 1 are comparable to the values obtained for NH$_3$/CO$_2$ ice mixtures with the highest value comparable to the value on silicate grains (Potapov, et al. 2019). The latter indicates comparable diffusion efficiencies of reactants on and within the dust, which is not surprising as in the present case we speak about diffusion along the pores of the dust.

To make the comparison between different experimental cases clearer, Figure 7 shows sketches and the main result (spectral indication of the ammonium carbamate formation) of three experiments, $CO_2$ + $NH_3$ ice mixtures (based on the results of previous studies (Bossa, et al. 2008; Noble, et al. 2014; Ghesquiere, et al. 2018; Potapov, et al. 2019; Potapov, et al. 2020b)), $CO_2$-ice/$NH_3$-ice layers and $CO_2$-ice/$MgSiO_3$-grains/$NH_3$-ice layers (based on the results of the present study).

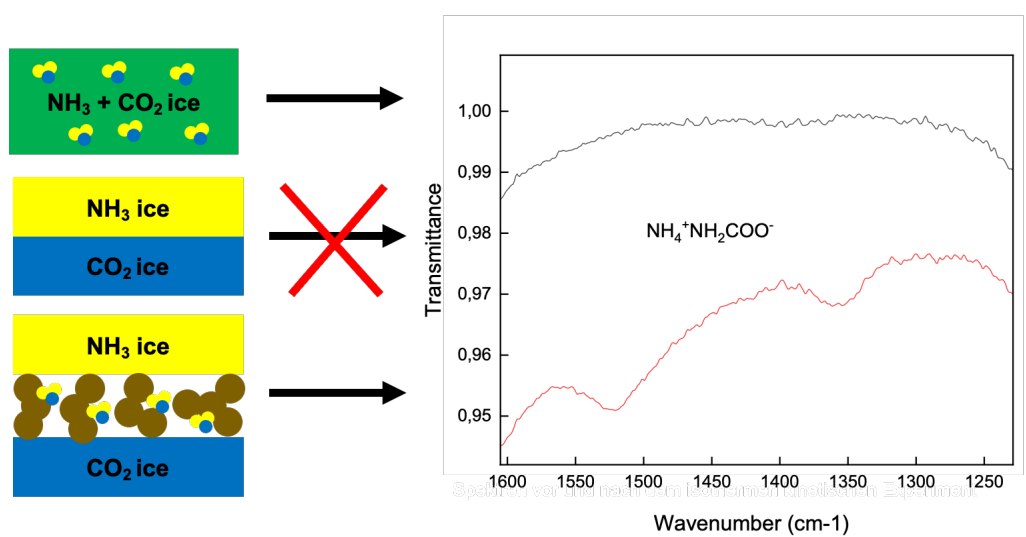

Figure 7. Sketches (left) of three experiments: $CO_2$ + $NH_3$ ice mixtures (top), $CO_2$-ice/$NH_3$-ice layers (middle), and $CO_2$-ice/$MgSiO_3$-grains/$NH_3$-ice layers (bottom); and the IR spectral indication of the ammonium carbamate formation (right) in the $CO_2$ + $NH_3$ ice mixtures and $CO_2$-ice/$MgSiO_3$-grains/$NH_3$-ice layers.

**Ammonium carbamate formation route**

The formation of carbamic acid ($NH_2COOH$) and ammonium carbamate from ammonia ($NH_3$) and carbon dioxide ($CO_2$) is well described in the literature and indeed has been extensively pursued as a potential carbon-capture vector (Bhown & Freeman 2011; Leung et al. 2014; Novek et al. 2016). Formation of carbamic acid, in the low $NH_3$ concentration limit, follows Scheme 1 (Figure 8a) and is found to have a substantial activation barrier of 48 kcal mol$^{-1}$ (Tsipis & Karipidis 2005). Formation of ammonium carbamate follows by proton exchange from carbamic acid to $NH_3$. At higher $NH_3$ concentrations, the reaction is catalyzed by the presence of a second $NH_3$ lowering the activation barrier to 32.1 kcal mol$^{-1}$ (Tsipis & Karipidis 2005) by stabilizing the charge on the carboxylate group and forming ammonium carbamate directly (Scheme 2, Figure 8b) rather than indirectly as in Scheme 1.

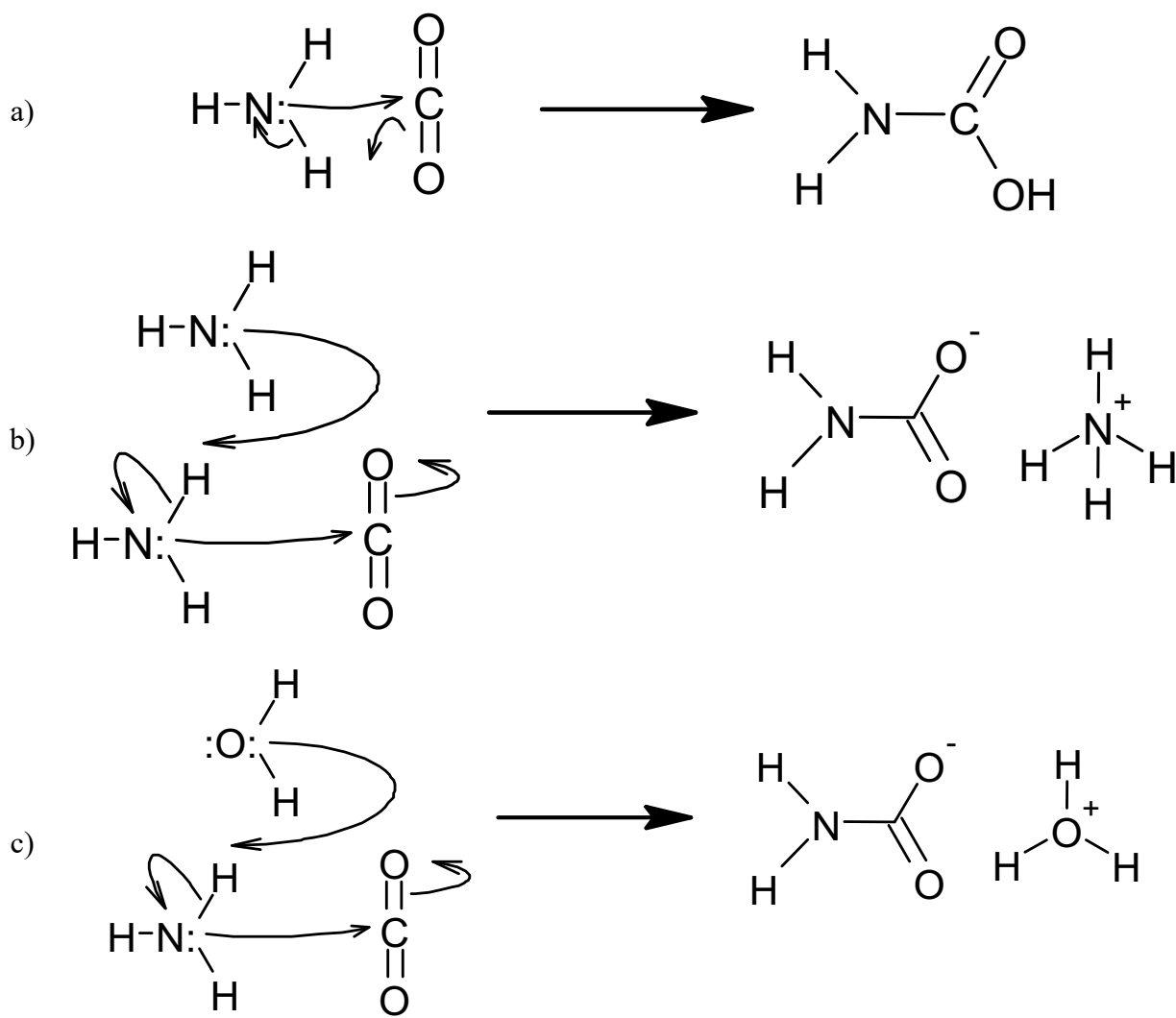

Figure 8. Carbamic acid and carbamate anion formation reaction schemes using the "curly arrow" notation common in mechanistic chemistry to identify the movement of electron pairs, and the making and breaking of chemical bonds. In (a), nucleophilic donation of the N atom lone pair on the $NH_3$ to the central C in $CO_2$ opens one of the C-O double bonds by transferring an electron pair from the bond to the terminal O atom. The enhanced electron density on the terminal O atom can then pull a proton from the in-coming $NH_3$ forming carbamic acid. In (b) and (c), the presence of a base ($NH_3$ and $H_2O$ respectively), the lone pairs on the corresponding N and O atoms in the base sees their nucleophilicity provide a pathway to the carbamate anion via proton transfer from the in-coming $NH_3$ to the base as the in-coming $NH_3$ donates its lone pair to the electrophilic carbon in the $CO_2$ forming the N-C bond as an electron pair transfers from a C-O double bond to a terminal O atom.

It is worth noting that proton transfer on ice surfaces has been extensively studied extensively in relation to stratospheric ozone depletion (Devlin et al. 1999; Devlin et al. 2002; Kondo et al. 2004b; Park & Kang 2005). Indeed, the analogous proton transfer reaction HCl + $NH_3$ → $NH_4^+Cl^-$ has also been explored on such surfaces (Kondo et al. 2004a). All this work points to proton transfer leading to acid ionisation or reaction on the surface and in the bulk of ice occurring at temperatures more than 80 K. The observation of reaction at and below 80 K on the surface of dust grain mimic materials that are known to have acid sites (Gierada et al. 2019) strongly suggests a catalytic role for the surface in this present study not inconsistent with the catalytic role of the $H_2O$ or second $NH_3$ entities in the mechanism in Figure 8. Furthermore, the earlier reports (Potapov, et al. 2019; Potapov, et al. 2020b) support this proposition. Thus, it is hard to deny that surface acid-base catalysis promotes ammonium carbamate formation on grain surfaces.

In our experiments, we do not detect carbamic acid. It was previously shown that carbamic acid can be observed in pure thermal $CO_2$ + $NH_3$ experiments only when $CO_2$ is present in an equal concentration with $NH_3$ (Noble, et al. 2014). This agrees with the catalytic role of a second $NH_3$ molecule. We propose that the mobility of both $CO_2$ and $NH_3$ should lead to diffusive meetings in which $CO_2$ can easily coordinate with more than one $NH_3$ molecule, thus producing ammonium carbamate. For $CO_2$ molecules that initially coordinate with a single $NH_3$ upon their first diffusive meeting, the higher barrier to reaction via Scheme 1 would likely avoid an immediate reaction, allowing either another $NH_3$ to arrive or for the $CO_2$ to diffuse again until an $NH_3$ pair was found, allowing the more favored Scheme 2 to proceed. The absence of carbamic acid in the experiments would therefore suggest that site-to-site diffusive hopping of either $CO_2$ or $NH_3$ at 80 K occurs on a timescale that is much shorter than the reaction via Scheme 1.

Water ($H_2O$) is also known to catalyze the reaction between $CO_2$ and $NH_3$ (Scheme 3, Figure 8c) and is found to lower the activation barrier even more to 29.7 kcal mol$^{-1}$ (Tsipis & Karipidis 2005). Though these barriers are substantial, proton tunnelling ensures that the low temperature solid state routes reported in the literature (e.g., (Molpeceres et al. 2021; Ferrero et al. 2024) and therein) are efficient. On the other hand, water molecules surrounding the reactants may block their diffusion. Influence of water on the formation efficiency of ammonium carbamate in $CO_2$-ice/$MgSiO_3$-grains/$NH_3$-ice systems will be a logical continuation of the present study. However, surface OH groups may not block diffusion but will catalytically promote carbamate formation chemistry.

The activation barriers provided above are for the gas phase. In the solid state, the barriers are much lower. For ammonium carbamate, it was estimated to be between 1 and 5 kJ mol$^{-1}$ (0.24 – 1.2 kcal mol$^{-1}$) depending on the surface (Noble, et al. 2014; Potapov, et al. 2020b). This reported barrier is consistent with the experimental 2 kJ mol$^{-1}$ barrier reported by Rosu-Finsen et al. for agglomeration of water on a silica surface (Rosu-Finsen et al. 2016) in the 15 to 25 K temperature range. Ammonia, like water, is a strongly hydrogen bonding species and we might expect similar diffusion kinetics to that of water under similar substrate and temperature conditions. Hence, we may reflect that at the temperatures at which the measurements were made, the reaction is diffusion limited.

In terms of the present work, the role of the dust grain surface is potentially two-fold and reflects the Langmuir-Hinshelwood (LH) chemistry occurring in this system. LH chemistry integrates diffusion and reaction on a surface. Firstly, the grain surface provides a platform for diffusive mixing of the spatially separated reactants on the dust grain surfaces. Then the surface provides sites where reactive encounters of the reactants occur. However, additionally, the surface of silicate minerals, such as used in the present work, often presents structures such as uncoordinated oxide and hydroxyl moieties. Such sites, like the secondary $NH_3$ and $H_2O$ in Schemes 2 and 3, open to accepting proton transfer as represented in Scheme 3; that is proton transfer specifically from the $NH_3$ group that forms the $NH_2$ moiety in the carbamate anion to the basic center on the surface. In this instance, at the low temperatures employed in the reported experiments, we can conclude that, in addition to being diffusion-limited, this reaction is surface-catalysed through a proton-transfer mechanism.

**Astrophysical Implication**

Dust extinction observations and dust models indicate a very efficient growth of nm-sized dust grains (see, e.g., (Draine 2003) for a review), first in diffuse and dense interstellar clouds and then in protoplanetary disks to aggregates of submicron-micron and submillimetre-millimetre sizes correspondingly. The formation of surface $NH_3$ may take place already in diffuse clouds, where it was detected in the gas phase, and both reactants certainly form in dense clouds, where $CO_2$ and $NH_3$ ices were detected (see, e.g., (Tielens 2013) for a review). However, the reaction leading to ammonium carbamate in mixed ices proceeds much more efficiently with an excess of $NH_3$ (Noble, et al. 2014), which is hard to imagine considering a much lower abundance of $NH_3$ as compared to $CO_2$ in the interstellar and circumstellar ices (Boogert, et al. 2015).

However, the following sequence of events may be considered as a means to explain efficient production of solid-phase ammonium carbamate: (i) the formation of $CO_2$ and $NH_3$ ices on the surfaces of dust grains in interstellar clouds; (ii) aggregation of these grains, providing porous dust structures and allowing a large internal surface to be created; and (iii) the heating of the aggregates in protostellar envelopes and protoplanetary disks that mobilizes the ices, allowing them to spread out (i.e. diffuse) into the internal surface and meet, leading to the formation of ammonium carbamate. Prerequisites for this scenario are exposure of the reactants to the bare grain surface and presence of the reactants on the same grain or on different grains that aggregate and have a common surface over which the reactants might diffuse. The scenario is reinforced by the recent first detection of ammonium carbamate in warm layers of the highly inclined disk d216-0939 by the James Webb Space Telescope (Potapov, et al. 2025a).

A question is whether this scenario can be applied to other reactions leading to the formation of other complex organic molecules. In many cases, solid-state reactants on astrophysical dust grains may not be mixed in high concentrations or in the required ratios; diffusion over the dust surface would then be a prerequisite for development of pathways to complexity in space. Thus, the present study opens a door to future important experimental and theoretical investigations.

## 5. Conclusions

The results of the study demonstrate efficient diffusion of $CO_2$ and $NH_3$ molecules over surfaces and through pores in layers of analogues of cosmic dust grains, leading to their reaction and formation of a complex organic molecule ammonium carbamate at 80 K, a temperature relevant to warm cosmic environments, such as protostellar envelopes and protoplanetary disks. Moreover, the presence of dust is a prerequisite for the reaction to proceed in cases where the reactants are separated (not mixed). The results point to the important role of dust grains in the formation of complex organic molecules and potentially efficient dust-promoted chemistry in warm astrophysical environments. It would be interesting and important to expand such experiments to other reactants, to see, first, whether the process of diffusion of molecules through the dust in common and, second, whether its efficiency depends on the molecule size. For instance, it would be important to the understand whether the chemical reactions involving large molecules can be triggered by their diffusion through the dust.

**Acknowledgments**

The present study was funded by the Deutsche Forschungsgemeinschaft (Heisenberg grant PO 1542/7-1).

**References**

Allamandola, L. J., Bernstein, M. P., Sandford, S. A., & Walker, R. L., Evolution of interstellar ices. 1999, Space Sci Rev, 90, 219

Bhown, A. S., & Freeman, B. C., Analysis and Status of Post-Combustion Carbon Dioxide Capture Technologies. 2011, Environ Sci Technol, 45, 8624, 10.1021/es104291d

Boogert, A. C. A., Gerakines, P. A., & Whittet, D. C. B., Observations of the Icy Universe. 2015, Annu Rev Astron Astr, 53, 541, https://doi.org/10.1146/annurev-astro-082214-122348

Bossa, J. B., Theule, P., Duvernay, F., Borget, F., & Chiavassa, T., Carbamic acid and carbamate formation in NH3:CO2 ices-UV irradiation versus thermal processes. 2008, Astron Astrophys, 492, 719, https://doi.org/10.1051/0004-6361:200810536

Burke, D. J., & Brown, W. A., Ice in space: surface science investigations of the thermal desorption of model interstellar ices on dust grain analogue surfaces. 2010, Phys Chem Chem Phys, 12, 5947

Congiu, E., Minissale, M., Baouche, S., Chaabouni, H., Moudens, A., Cazaux, S., Manico, G., Pirronello, V., & Dulieu, F., Efficient diffusive mechanisms of O atoms at very low temperatures on surfaces of astrophysical interest. 2014, Faraday Discuss, 168, 151, 10.1039/c4fd00002a

d'Hendecourt, L. B., & Allamandola, L. J., Time-Dependent Chemistry in Dense Molecular Clouds .3. Infrared Band Cross-Sections of Molecules in the Solid-State at 10-K. 1986, Astronomy & Astrophysics Supplement Series, 64, 453

Devlin, J. P., Uras, N., Rahman, M., & Buch, V., Covalent and ionic stales of strong acids at the ice surface. 1999, Israel J Chem, 39, 261, 10.1002/ijch.199900033

Devlin, J. P., Uras, N., Sadlej, J., & Buch, V., Discrete stages in the solvation and ionization of hydrogen chloride adsorbed on ice particles. 2002, Nature, 417, 269, 10.1038/417269a

Draine, B. T., Interstellar dust grains. 2003, Annu Rev Astron Astr, 41, 241

Ehrenfreund, P., Boogert, A., Gerakines, P., & Tielens, A., Apolar ices. 1998, Faraday Discuss, 109, 463, DOI 10.1039/a800608c

Ferrero, S., Ceccarelli, C., Ugliengo, P., Sodupe, M., & Rimola, A., Formation of Interstellar Complex Organic Molecules on Water-rich Ices Triggered by Atomic Carbon Freezing. 2024, Astrophys J, 960, 22, 10.3847/1538-4357/ad0547

Gerakines, P. A., Schutte, W. A., Greenberg, J. M., & van Dishoeck, E. F., The Infrared Band Strengths of H2o, Co and Co2 in Laboratory Simulations of Astrophysical Ice Mixtures. 1995, Astron Astrophys, 296, 810

Ghesquiere, P., Ivlev, A., Noble, J. A., & Theule, P., Reactivity in interstellar ice analogs: role of the structural evolution. 2018, Astron Astrophys, 614, A107

Ghesquiere, P., Mineva, T., Talbi, D., Theule, P., Noble, J. A., & Chiavassa, T., Diffusion of molecules in the bulk of a low density amorphous ice from molecular dynamics simulations. 2015, Phys Chem Chem Phys, 17, 11455, 10.1039/c5cp00558b

Gierada, M., De Proft, F., Sulpizi, M., & Tielens, F., Understanding the Acidic Properties of the Amorphous Hydroxylated Silica Surface. 2019, J Phys Chem C, 123, 17343, 10.1021/acs.jpcc.9b04137

Hudgins, D. M., Sandford, S. A., Allamandola, L. J., & Tielens, A. G. G. M., Midinfrared and Far-Infrared Spectroscopy of Ices - Optical-Constants and Integrated Absorbances. 1993, Astrophysical Journal Supplement Series, 86, 713

Kondo, M., Kawanowa, H., Gotoh, Y., & Souda, R., Hydration of the HCl and NH3 molecules adsorbed on amorphous water-ice surface. 2004a, Appl Surf Sci, 237, 509, 10.1016/j.apsusc.2004.06.063

Kondo, M., Kawanowa, H., Gotoh, Y., & Souda, R., Ionization and solvation of HCl adsorbed on the D2O-ice surface. 2004b, J Chem Phys, 121, 8589, 10.1063/1.1804153

Kouchi, A., Furuya, K., Hama, T., Chigai, T., Kozasa, T., & Watanabe, N., Direct Measurements of Activation Energies for Surface Diffusion of CO and CO on Amorphous Solid Water Using In Situ Transmission Electron Microscopy. 2020, Astrophys J Lett, 891, L22, 10.3847/2041-8213/ab78a2

Leung, D. Y. C., Caramanna, G., & Maroto-Valer, M. M., An overview of current status of carbon dioxide capture and storage technologies. 2014, Renew Sust Energ Rev, 39, 426, 10.1016/j.rser.2014.07.093

Marchione, D., Rosu-Finsen, A., Taj, S., Lasne, J., Abdulgalil, A. G. M., Thrower, J. D., Frankland, V. L., Collings, M. P., & McCoustra, M. R. S., Surface Science Investigations of Icy Mantle Growth on Interstellar Dust Grains in Cooling Environments. 2019, Acs Earth Space Chem, 3, 1915, 10.1021/acsearthspacechem.9b00052

Mispelaer, F., et al., Diffusion measurements of CO, HNCO, H2CO, and NH3 in amorphous water ice. 2013, Astron Astrophys, 555, A13

Molpeceres, G., Kastner, J., Fedoseev, G., Qasim, D., Schomig, R., Linnartz, H., & Lamberts, T., Carbon Atom Reactivity with Amorphous Solid Water: H2O-Catalyzed Formation of H2CO. 2021, J Phys Chem Lett, 12, 10854, 10.1021/acs.jpclett.1c02760

Noble, J. A., Theule, P., Duvernay, F., Danger, G., Chiavassa, T., Ghesquiere, P., Mineva, T., & Talbi, D., Kinetics of the NH3 and CO2 solid-state reaction at low temperature. 2014, Phys Chem Chem Phys, 16, 23604

Novek, E. J., Shaulsky, E., Fishman, Z. S., Pfefferle, L. D., & Elimelech, M., Low-Temperature Carbon Capture Using Aqueous Ammonia and Organic Solvents. 2016, Environ Sci Tech Let, 3, 291, 10.1021/acs.estlett.6b00253

Park, S. C., & Kang, H., Adsorption, ionization, and migration of hydrogen chloride on ice films at temperatures between 100 and 140 K. 2005, J Phys Chem B, 109, 5124, 10.1021/jp045861z

Potapov, A., Bouwman, J., Jäger, C., & Henning, T., Dust/ice mixing in cold regions and solid-state water in the diffuse interstellar medium. 2021, Nat Astron, 5, 78, https://doi.org/10.1038/s41550-020-01214-x

Potapov, A., Jäger, C., & Henning, T., Ice coverage of dust grains in cold astrophysical environments. 2020a, Phys Rev Lett, 124, 221103, https://doi.org/10.1103/PhysRevLett.124.221103

Potapov, A., Jäger, C., & Henning, T., Thermal formation of ammonium carbamate on the surface of laboratory analogs of carbonaceous grains in protostellar envelopes and planet-forming disks. 2020b, ApJ, 894, 110, 10.3847/1538-4357/ab86b5

Potapov, A., Jäger, C., Mutschke, H., & Henning, T., Trapped Water on Silicates in the Laboratory and in Astrophysical Environments. 2024, Astrophys J, 965, 48, 10.3847/1538-4357/ad2c07

Potapov, A., Linz, H., Bouwman, J., Rocha, W., Martin, J., Wolf, S., Henning, T., & Terada, H., Simple molecules and complex chemistry in a protoplanetary disk. A JWST investigation of the highly inclined disk d216-0939. 2025a, A&A, 97, A53, https://doi.org/10.1051/0004-6361/202453385

Potapov, A., & McCoustra, M. R. S., Physics and chemistry on the surface of cosmic dust grains: a laboratory view. 2021, Int Rev Phys Chem, 40, 299, https://doi.org/10.1080/0144235X.2021.1918498

Potapov, A., McCoustra, M. R. S., Tazaki, R., Bergin, E., Bromley, S., Garrod, R. T., & Rimola, A., Is cosmic dust porous? 2025b, The Astronomy and Astrophysics Review, accepted for publication, https://arxiv.org/abs/2509.10292

Potapov, A., Semenov, D., Jäger, C., & Henning, T., Formation of CO2 driven by photochemistry of water ice mixed with carbon grains 2023, ApJ, 954, 167, 10.3847/1538-4357/acebcc

Potapov, A., Theule, P., Jäger, C., & Henning, T., Evidence of surface catalytic effect on cosmic dust grain analogues: the ammonia and carbon dioxide surface reaction. 2019, ApJL, 878, L20, 10.3847/2041-8213/ab2538

Rosu-Finsen, A., Marchione, D., Salter, T. L., Stubbing, J. W., Brown, W. A., & McCoustra, M. R. S., Peeling the astronomical onion. 2016, Phys Chem Chem Phys, 18, 31930, 10.1039/c6cp05751a

Tielens, A. G. G. M., The molecular universe. 2013, Reviews of Modern Physics, 85, 1021

Tsipis, C. A., & Karipidis, P. A., Mechanistic insights into the Bazarov synthesis of urea from NH3 and CO2 using electronic structure calculation methods. 2005, J Phys Chem A, 109, 8560, 10.1021/jp051334j

Tsuge, M., Molpeceres, G., Aikawa, Y., & Watanabe, N., Surface diffusion of carbon atoms as a driver of interstellar organic chemistry. 2023, Nat Astron, 7, 1351, 10.1038/s41550-023-02071-0